\documentclass[10pt]{article}
\usepackage{bbm}
\usepackage[final]{graphics}
\usepackage{amsmath}
\usepackage{amsfonts,amsbsy}
\usepackage{amssymb}
\usepackage{graphicx}

\def\empile#1\over#2{\mathrel{\mathop{\kern 0pt#1}\limits_{#2}}}

\newcommand{\slcalP}{\raise.15ex\hbox{$/$}\kern-.63em\hbox{$\cal P$}}

\catcode`\@=11

\newcount\@tempcntc
\def\@citex[#1]#2{\if@filesw\immediate\write\@auxout{\string\citation{#2}}\fi
  \@tempcnta\z@\@tempcntb\m@ne\def\@citea{}\@cite{%
        \@for\@citeb:=#2\do%
    {\@ifundefined{b@\@citeb}%
        {\@citeo\@tempcntb\m@ne\@citea%
                \def\@citea{,\penalty\@m\ }{\bf ?}\@warning%
                {Citation `\@citeb' on page \thepage \space undefined}}%
        {\setbox\z@\hbox{\global\@tempcntc0\csname b@\@citeb\endcsname\relax}%%
     \ifnum\@tempcntc=\z@ \@citeo\@tempcntb\m@ne%
       \@citea\def\@citea{,\penalty\@m}%
       \hbox{\csname b@\@citeb\endcsname}%
     \else%
      \advance\@tempcntb\@ne%
      \ifnum\@tempcntb=\@tempcntc%
      \else\advance\@tempcntb\m@ne\@citeo%
      \@tempcnta\@tempcntc\@tempcntb\@tempcntc\fi\fi}}\@citeo}{#1}}%
 
\def\@citeo{\ifnum\@tempcnta>\@tempcntb\else\@citea
  \def\@citea{,\penalty\@m}%
  \ifnum\@tempcnta=\@tempcntb\the\@tempcnta\else
   {\advance\@tempcnta\@ne\ifnum\@tempcnta=\@tempcntb \else
\def\@citea{--}\fi
    \advance\@tempcnta\m@ne\the\@tempcnta\@citea\the\@tempcntb}\fi\fi}

\catcode`\@=12

\begin{document}

\title{\bf  From AGS-SPS and Onwards to the LHC}
\author{Larry McLerran}
\maketitle
\begin{center}
 Physics Department and Riken Brookhaven Research Center, Building 510A\\ Brookhaven National Laboratory,
  Upton, NY-11973, USA
\end{center}

\begin{abstract}
 I review the history of the efforts using heavy ion collisions to make new forms of
 matter.  I discuss both the development of the theoretical ideas about such new forms of
 matter, and as well the past, present and planned experimental efforts.  I also highlight the development
 of this activity in both India and China. 
\end{abstract}

\section{Origins of the Study of Ultra-Relativistic Heavy Ion Collisions}

The first meeting where the idea of using the collisions of ultra-relativistic heavy ions to make and study
new forms of matter was {\it Worskhop on BeV Collisions of Heavy Ions: How and Why}, which was held in Bear Mountain New York, Nov. 29 - Dec 1974.\cite{bearmountain}  This meeting marked the beginning of our field and as such deserves special emphasis.  

The introduction and summary of that meeting was written by Leon Lederman and Joseph Weneser:
{\bf 
\begin{center} 
Introduction and Summary: \\
\end{center}
The history of physics teaches us that profound revolutions arise from a gradual perception that certain observations can be accommodated only by radical departures from current thinking.  The workshop addressed itself to the intriguing question of the possible existence of a nuclear world quite different from the one we have learned to accept as familiar and stable.}

Professor T. D. Lee of Columbia University played a central role in this meeting and in initiating this scientific endeavor and I quote from his statement at the meeting:

{\bf
It would be interesting to explore new pheomena by distributing high energy or high energy density over a relatively large volume.\\ 
}

\begin{figure}[htbp]
\begin{center}
\resizebox*{!}{6cm}{\includegraphics{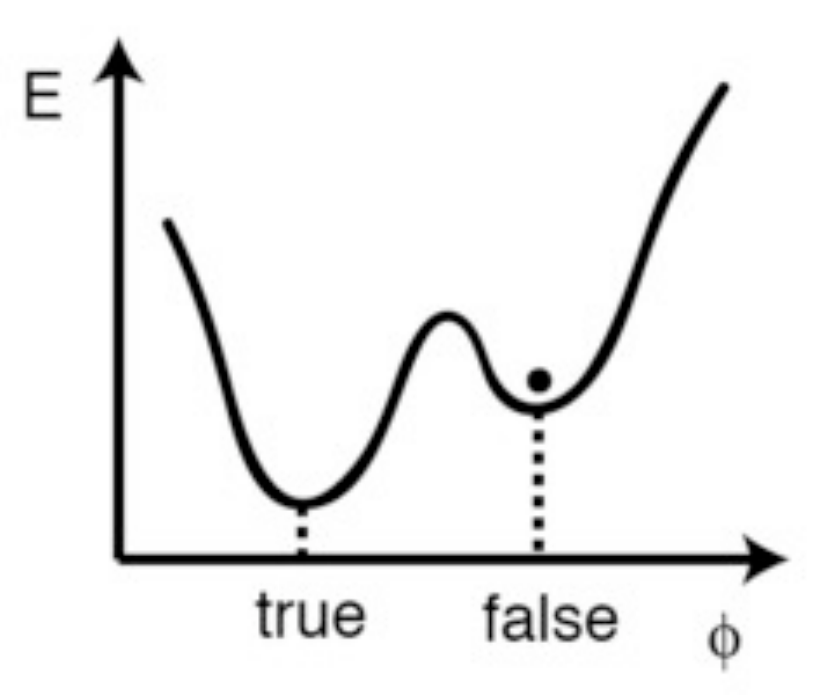}}
\end{center}
\caption{The energy of the vacuum as a function of the expectation value of a scalar field.
}
\label{leewick}
\end{figure}

At this time, Professor Lee had formulated his idea about Lee Wick matter.\cite{leewick}  The idea
was based on a scalar field theory model, and argued that we might currently live in an unstable vacuum.  This is shown in Figure 1.  As a function of the expectation value of a scalar field, the energy may have multiple minima.  By an accident of history, we might not live in the lowest energy state.  If in a heavy ion collision, we would heat the system we might make a transition to the true vacuum.\cite{voloshin}  If such a transition occurred,  it would of course be catastrophic.  Fortunately, such a scenario may be absolutely ruled out for heavy ion collisions because high energy cosmic ray interactions would have long ago triggered such a cataclysm, if it were possible.  It nevertheless showed the intrinsic merit in studies  of the properties of the vacuum in which we live, and the nontrivial physics which might be learned should we heat up the vacuum.  It also provided the intellectual seeds of inflationary cosmology, which have reshaped the way we think about the big bang.\cite{guth}

In this talk, I will discuss the development of ideas associated with very high energy density matter, and the search for such matter in the collisions of ultra-relativistic heavy ion collisions.  This talk will be historical in nature, and as such, I begin wih a quote from Mark Twain:

{\bf  Herodotus says: "Very few things happen at the right time, and the rest do not happen at all:  The conscientious historian will correct these defects"}

I will of course try to make this talk an accurate representation of the events, but of course time and personal taste distorts the picture.  I apologize that limitations of space often make it impossible for me to refer to original literature, and instead I direct the reader to relevant reviews.
 
I will focus on several issues:

\begin{itemize}
\item{The Concept of the Quark Gluon Plasma}
\item{Properties of the Quark Gluon Plasma}
\item{The Space-Time Structure of Hadronic Collisions}
\item{The Color Glass Condensate and Glasma}
\item{Theory and Experiment}
\end{itemize}

Physics is a field which combines together flights of fantasy which are occasionally connected together
by the language of mathematics, and sometimes, but rarely, have a realization in nature. These ideas arise sometimes by a discovery by direct observation of nature, or sometimes by an insight which is based on pure thought.  I want to try to capture some of the intellectual excitement of scientific discovery as partially formed ideas realize themselves, and over time become part of our scientific knowledge. 

The talk as presented at the meeting has many photos and illustrations which cannot be presented 
in a conference proceedings.  An original powerpoint version of the talk is available at the
{\bf //www.larrymclerran.com/Site/Talks.html}.

\section{Early Work on the Phase Diagram of QCD} 

The first time to my knowledge that Quark Matter was proposed in the scientific literature was a model for neutron stars composed of heavy quarks by N. Itoh.\cite{itoh}  A little later,
Pete Carruthers provided theoretical arguments for the possible existence of such matter,\cite{carruthers} arguments which became quite compelling with the later work of Cabibbo and Parisi\cite{cabibbo} and 
Collins and Perry.\cite{collins}  The arguments of Collins and Perry used the asymptotic freedom of strong interactions to argue that at extremely high baryon number density, matter would form a gas of weakly interacting quarks.  Baym and Chin later computed the first order terms due to interactions,\cite{baym} and Freedman and Mclerran computed the three loop contributions to the thermodynamic potential.\cite{freedman}  The latter computation was the first done to high enough order so that one could explicitly see the density dependence of the interaction strength, and the cancellation of infrared divergences important for a finite result. A short time later, the first computations at finite temperature were done by Shuryak,\cite{shuryak} and  by Kapusta.\cite{kapusta}  Ed Shuryak coined the expression Quark Gluon Plasma to describe the gas of very high energy density quarks and gluons.

\begin{figure}[htbp]
\begin{center}
\begin{tabular}{l l l}
\includegraphics[width=0.40\textwidth]{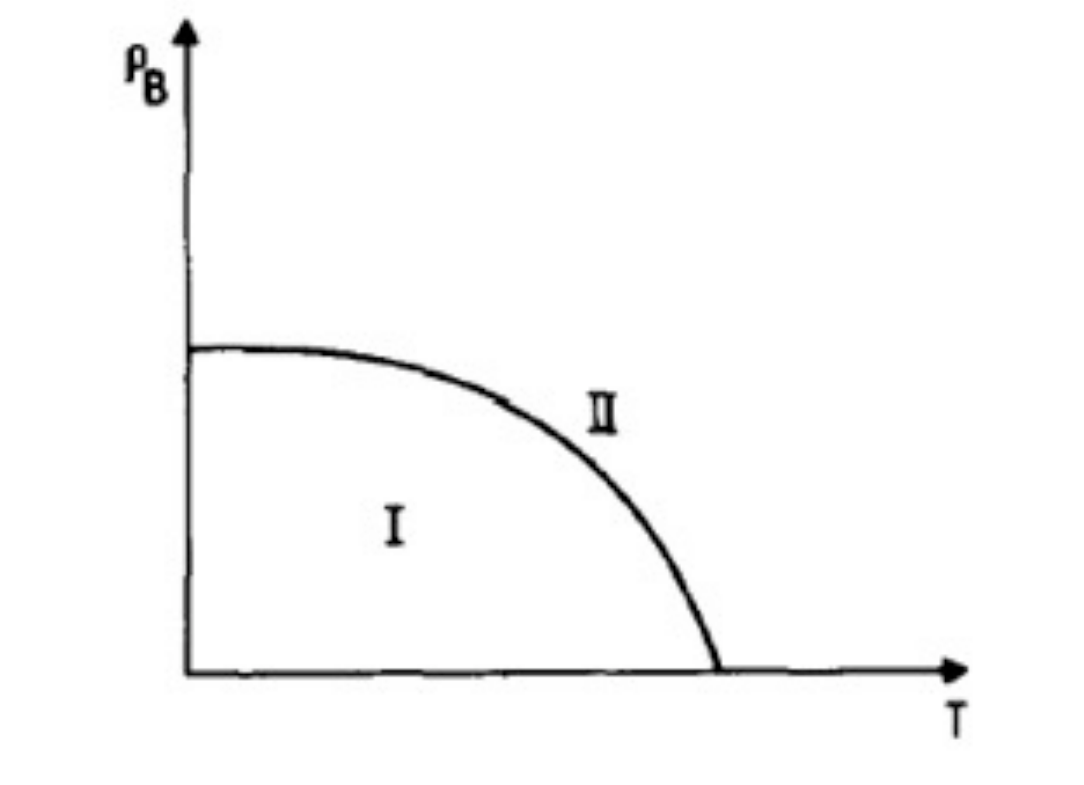} & & 
\includegraphics[width=0.60\textwidth]{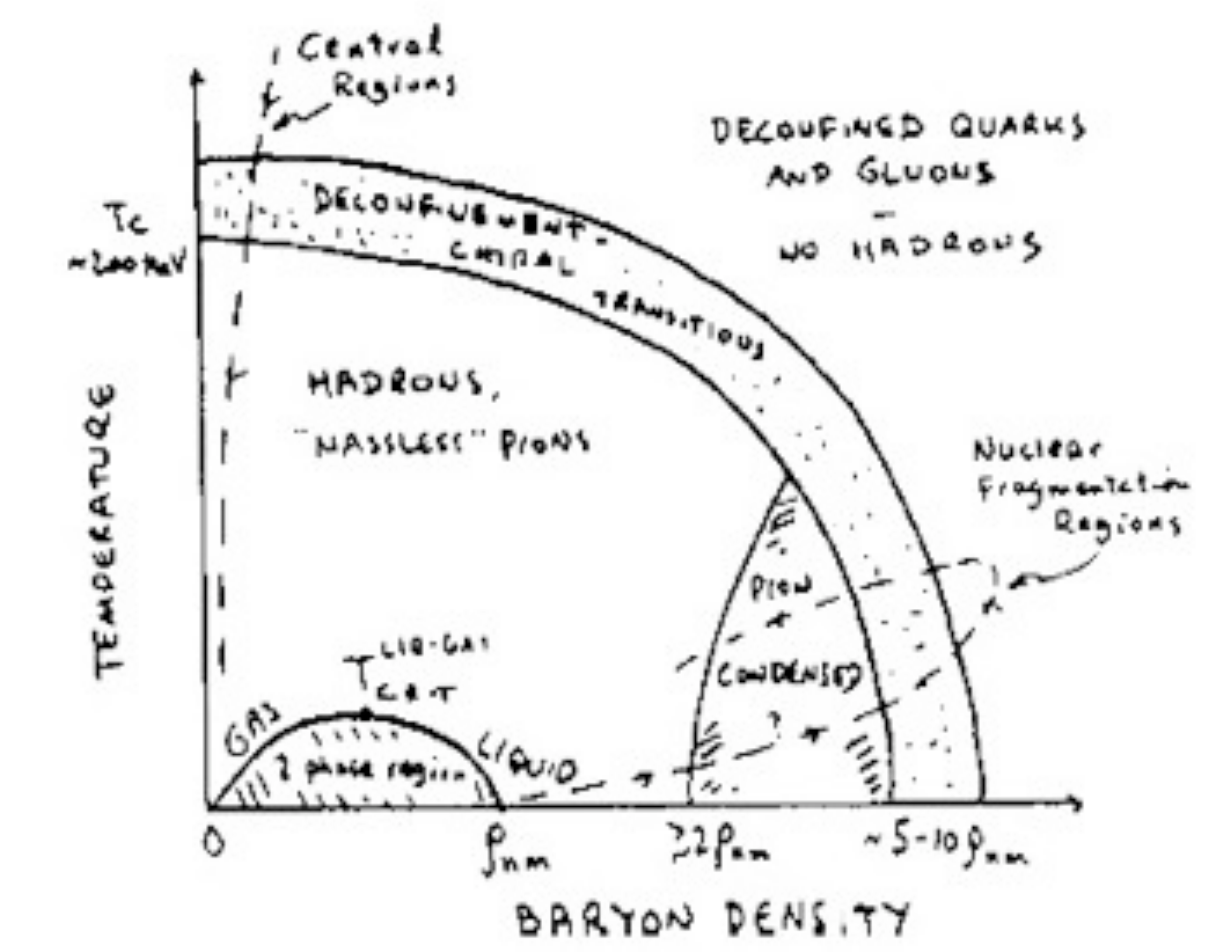}\\
 & & \\
 a & & b \\
\end{tabular}
\end{center}
\caption{(a) The "phase diagram" of QCD presented as envisaged by Cabibbo and Paris. (b) The phase diagram of QCD presented in the US 1983 NSAC Long Range Plan meeting .
}
\label{cabibbo}
\end{figure}

The basic idea for the phase diagram of QCD was envisioned in the paper by Cabibbo and Parisi,\cite{cabibbo} and is shown in Fig. \ref{cabibbo}a.  By 1983,  Gordon Baym had understood basicaly how heavy ion collisions might scan through this phase diagram, and included in his now famous plot, the possibility of pion condensation and the liquid gas phase transitions, as shown in Fig. \ref{cabibbo}b.

In 1980, Helmut Satz organized the meeting Statistical Mechanics of Quarks and Hadrons in Bielefeld, Germany,\cite{bielefeld} and brought together a wide variety of  theorists interested in the properties of matter at very high energy density.  (This and the meeting at Berkeley in 1978 were the precursors of the Quark Matter series of meetings.\cite{darmstadt} ) Attendees included T. D. Lee, "Papa" Migdal, Gordon Baym and a young Andrei Linde.  There was excitement about the recent results of Polyakov and Susskind concerning the
confinement phases transition in strong coupling in lattice gauge theory.
This groundbreaking meeting introduced the first results from lattice gauge theory
computations at finite temperature,\cite{svetitsky}-\cite{kuti} which showed clear evidence of a conifnement-deconfinement phase transition.  Rob Pisarksi presented his results with David Gross and Larry Yaffe showing how to include instantons into finite temperature computations.\cite{gpy}  For many of us, this meeting shaped the remainder of our careers.  It was packed with the intellectual excitement
of newly forming ideas.  We made strong friendships which have lasted the almost thirty years since this meeting.

Helmut Satz and his young collaborators were setting up to do lattice gauge theory computations at finite temperature as the meeting took place,\cite{engels} and subsequently developed a successful  program of systematic study.  This program continues now\cite{karsch}, and at Brookhaven National Laboratory, we have a strong effort headed by Fritjof Karsch who was at that time a young graduate student of Helmut. 

As time has evolved, a variety of methods, computational and analytic have given us an increasingly rich 
picture of the phase diagram of QCD, Fig. \ref{history}.
\begin{figure}[htbp]
\begin{center}
\resizebox*{!}{6cm}{\includegraphics{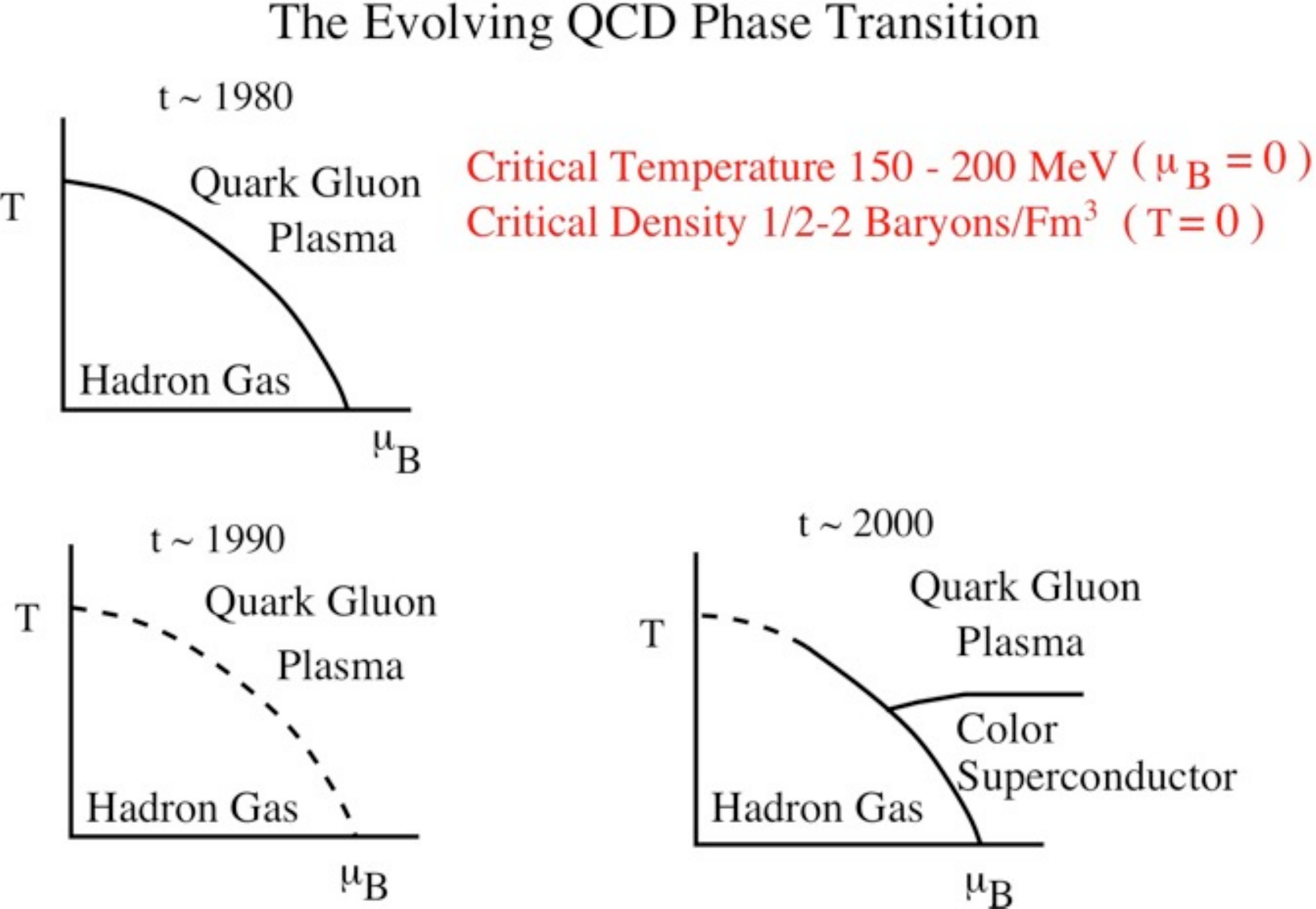}}
\end{center}
\caption{The evolving phase diagram of QCD.
}
\label{history}
\end{figure}
I drew this figure almost ten years ago, to show that theoretical conceptions, in the absence of hard scientific fact, evolve.  At about the time of the Bielefeld meeting, it was believed that there was a simple line in the
baryon chemical potential - temperature plane which separated the confined and deconfined world with
a phase transition.  By 1990, many people believed that there was only a cross over, where the energy
density changed quickly, but no real phase transition between the hadronic world and that of the quark gluon plasma.  By the year 2000, it had all changed again, with the possibility of a line of first
order phase transitions, ending in a critical point, and color superconductivity at very high density.\cite{alford}  

I think there has been more change in the interim since this figure was first drawn.  There is likely
an entirely new phase of matter at low temperature and intermediate baryon density, which is called the
quarkyonic phase.\cite{pisarski1}.  This phase has confined quarks, even though for bulk quantities
like the energy density, the dependence is as if there were free quarks.  The idea is that the interactions
are perturbative inside the Fermi sea, but there is confinement on the Fermi surface.  This phase
cannot have color superconductivity, and the issue of chiral symmetry breaking or restoration is not yet understood.
The various phases may be categorized at large numbers of colors, $N_c$.  In this limit there is an order parameter which corresponds to confinement, and an order parameter corresponding to the baryon number density.  The phases are the hadronic phase which has zero baryon number density and is confined, the deconfined phase which has no confinement and finite baryon number density,
and the quarkyonic phase which has confinement and finite baryon number density.  These phases have energy densities which scale with the number of colors like $O(1)$, $N_c^2$, and $N_c$ respectively.

The properties of QCD matter have been much studied in perturbation theory.  At finite temperature,
Linde proved there were finite number of orders of perturbation theory which could be computed before the series broke down.\cite{linde}.  The finite temperature computations are largely completed.\cite{perttheory}.  They yield a series which is poorly convergent.  With the development of the hard thermal loop Lagranegan of Pisarski and Braaten,\cite{htl}, one has found ways of re-summing the series so that it is superficially more convergent.\cite{iancu}  It may be that the Quark Gluon Plasma is at current energy scales, intrinsically strongly coupled, and methods based on the AdSCFT correspondence might have some use.\cite{adscft}

\section{Space Time Picture of Heavy Ion Collisions}

To make a Quark Gluon Plasma in a laboratory requires the collisions of hadrons at very high energy.  There were several frameworks developed for understanding such collisions.  Landau long ago formulated a hydrodynamic description of high energy collisions, but it fell to disfavor after the
discovery of the aysmptotc freedom of high energy intereactions.  Landau hydrodynamical theory requires that the colliding particles stop in the center of mass frame, which is impossible if high energy particles have perturbative interactions at high momentum transfer. Bjorken and Feynman.\cite{bjfeynman} developed
a description consistent with our understanding of strong interactions., Within this framework, there were early estimates of the energy densities achievable in high energy collisions,\cite{shuryakanishetty}  Bjorken developed a hydrodynamical model for the description of such collisions at a time later than some initial time at which the matter thermalized.\cite{bjorken}  This picture forms the basis of most modern hydrodynamical studies.\cite{modernhydro}

\begin{figure}[htbp]
\begin{center}
\includegraphics[width=0.70\textwidth]{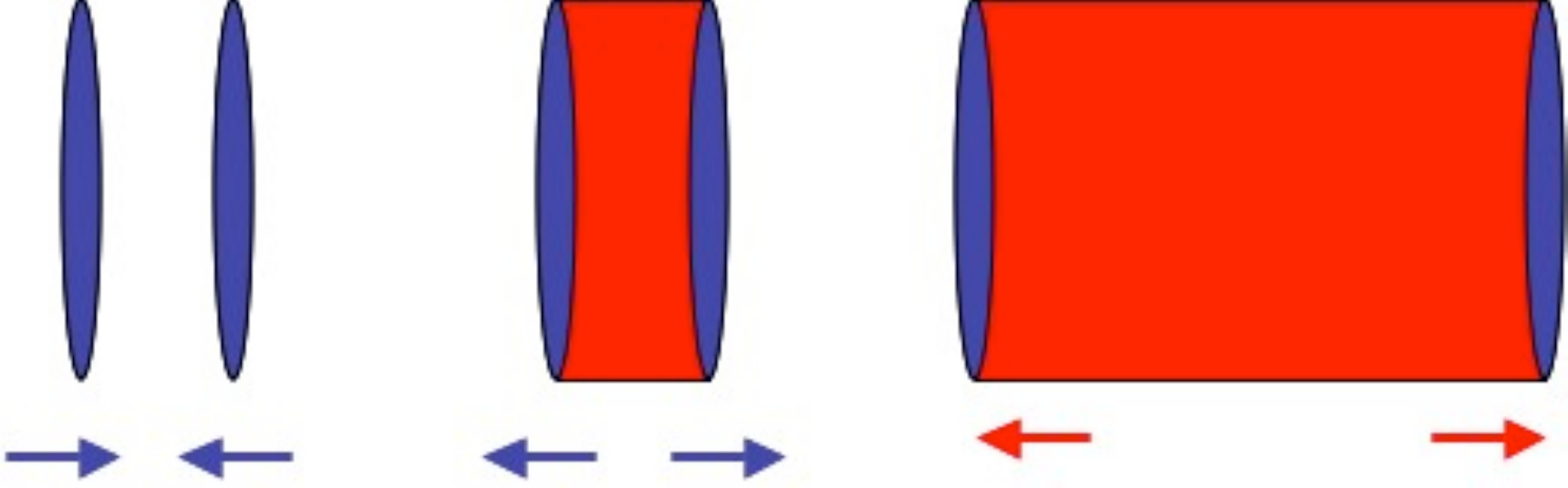}
\end{center}
\caption{A picture of high energy collisions in the Bjorken description.  The blue represents 
the valence quarks and red the produced hot matter.
}
\label{history}
\end{figure}
The hydrodynamic studies until very recently have assumed perfect fluid dynamics, that is fluid dynamics with zero viscosity.  Such  hydrodynamic computations are very successful in describing momentum space distributions of produced particle and collective effects such as flow. The space time evolution of particles produced in 
these collisions provide input for the computation of many observables.  There is some uncertainty in the 
in initial collisions for the computation of flow observables, which can be absorbed into uncertainties in the viscosity of the produced matter.  The success of these computations has led to the paradigm of the 
Strong Quark Gluon Plasma, and has led to speculations as to the nature of the validity of the small
viscosity limit.\cite{son}

\section{The Color Glass Condensate and Glasma}

\begin{figure}[htbp]
\begin{center}
\begin{tabular}{l l l}
\includegraphics[width=0.40\textwidth]{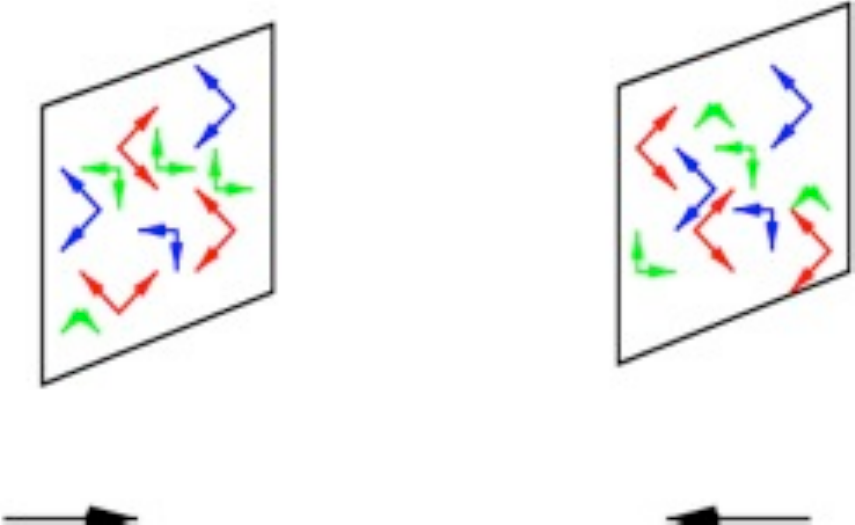} & & 
\includegraphics[width=0.50\textwidth]{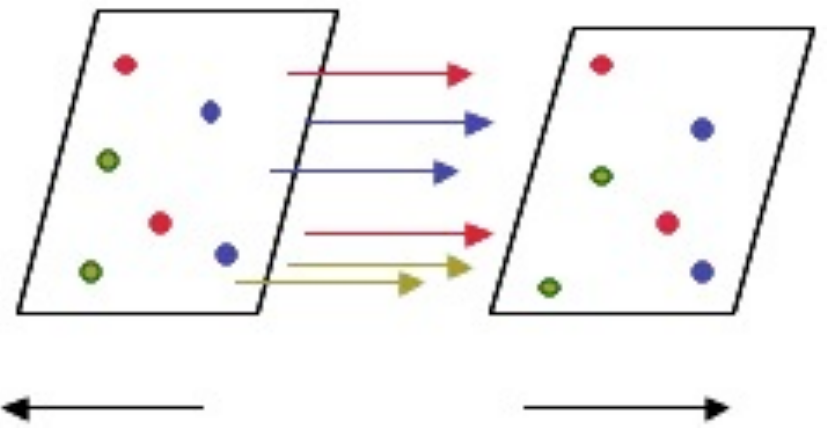}\\
 & & \\
 a & & b \\
\end{tabular}
\end{center}
\caption{(a) Two sheets of Color Glass Condensate colliding.  (b) The Glasma immediately after a high energy collision.
}
\label{cgc}
\end{figure}

The Color Glass Condensate is the matter associated with the wee partons of a high energy
hadronic wavefunction.\cite{mv}-\cite{iancucgc}  This matter has very high energy density.  It has properties similar to Bose condensates and to spin glasses.  It was motivated based on observations
in deep inelastic scattering.  The idea is that as one goes to higher and higher energies, there are more
and more gluons in the hadron wavefunction.  They have to squeeze together, and highly occupy phase space, making a highly coherent high density system of gluons.  This matter controls the high energy
limit of hadronic scattering and provides the initial conditions for the matter made in such collisions.
A picture of sheets of Color Glass Condensate colliding is shown in Fig. \ref{cgc}a.

The Glasma is the matter made after two sheets of Colored Glass collide.\cite{weigert}-\cite{lappi}  The gluons in the Color Glass Condensate are transversely polarized relative to the collision axis.  Immediately after the collision,
these fields change into longitudinal color electric and magnetic fields, as shown in Fig. \ref{cgc}b.  These fields carry Chern-Simons number, and generate long range correlation in rapidity.  They arise from the charging of the
nuclei as they pass through one another with color electric and color magnetic charge.  It is this matter which eventually  forms the Quark Gluon Plasma.

The theoretical ideas concerning these two forms of matter have much of their origin in the RHIC heavy ion collisions, and will presumably be tested at RHIC and LHC.

\section{Early Days of Heavy Ion Collisions:  The Bevalac, AGS and SPS}

The pioneering work on ultra-relativistic heavy ion collisions at the Bevalac provided the seeds both intellectual and spiritual of the efforts which were to follow at the SPS and RHIC.  There were two major
experiments, the Plastic Ball and the Streamer Chamber.  The Frankfurt theory group, most notably Stocker and Greiner, argued strongly for hydrodynamic like flow in such collisions.\cite{flowtheory}  This was observed in
collisions gold nuclei as a function of the centrality of the collision and the beam energy.\cite{flowexperiment} 

 I was privileged to see the new developments at the Bevalac  on visits to LBL as a young man, and  the genuine excitement and messianic enthusiasm of young experimentalists such as Rienhard Stock, Hans Gutbrod, Hans-Ake Gustafson, Art Poskanzer and Han Georg Ritter,
and young theorists such as Miklos Gyulassy and Hoerst Stocker.   

The AGS experiments studied the high baryon density region of ultrarelativistic nuclear collisions.\cite{reviewags}
One of the questions addressed was when would there be an approximately baryon number free central region.  There was an early analysis by Busza and Goldhaber which indicated that this would need very high energyes, and this analysis was confirmed in the experiments.\cite{busza}
In addition, there were studies of the flavor composition of high energy density matter, and confirmation that statisiiical models indeed described the data.\cite{statmodels}

The SPS program involved the experiments NA44, NA45/CERES, NA49, NA50, NA52/NEWMASS, WA97/NA57, WA98 and NA60.  It was a massive effort, with involvement of worldwide experimental groups. In Feb. of 2002, Maurice Jacob and Ulrich Heinz presented an assesment of the results of the SPS in {\it "Evidence for a New State of Matter: An Assessment of the Results from the CERN Lead Beam Program"}\cite{heinz}  I quote from the document:

{\bf A common assessment of the collected data leads us to conclude that we now have compelling evidence that a new state of matter has indeed been created, at energy densities which have never been reached over appreciable volumes in laboratory experiments before and which exceed by more than a factor of 20 that of normal nuclear matter.  The new state of matter found in heavy ion collisions at the SPS features many of the characteristics of the theoretically predicted Quark Gluon Plasma}

The evidence put forward to support the case was 
\begin{itemize}
\item{Bose Einstein Interferometry:  The time and size scales of the collision
for the collisions were much larger than natural time and size scales.}
\item{Flavor Abundances:  Ratios of the total numbers of produced hadrons of different flavors  was
consistent with a statistical model as had been the case at the AGS}
\item{Rho Melting:  The Rho meson was much broader than in vacuum,.  This most likely arose from multiple scattering in a hadronic medium}
\item{$J/\Psi$ Melting:  The $J/\Psi$ resonance was produced much less abundantly than
would be expected from ordinary $pp$ interactions. }
\end{itemize}

I believe that the claim that matter was made which exceeded by a factor of $20$ that of nuclear matter is supported by the SPS experiments.  The evidence that the properties of this matter was consistent with a Quark Gluon Plasma is modestly and weakly stated  in the document, and is not a statement of discovery of a Quark Gluon Plasma.  The $J/\Psi$ melting, which at the time was argued to be  the strongest evidence for a Quark Gluon plasma, in light of the RHIC experiments, has proven more difficult to interpret.\cite{heinz}-\cite{expwhitepaper}  Another problem for the SPS experiments was that it was not possible to simultaneously describe the transverse momentum distributions of partilces
and elliptic flow within a hydrodynamical model.  This indicates that the matter produced may not have been well thermalized.  Nevertheless, we all recognize the importance of the seminal achievments of this effort.  

\section{The RHIC Program}

It is impossible to due justice to the beautiful  results from the RHIC program in a talk of this scope.  The program is huge, and involves about 1200 physicists from over 50 countries. There have been over 2000 publications, many of which are summarized in the whitepapers from the various experiments.\cite{expwhitepaper}  By any measure of scientific quantity, quality interest  and excitement, the program has been both successful and rich in its impact.

Among the discoveries at RHIC are:
\begin{itemize}  

\item{ Remarkably large elliptic flow in the collisions.}
\item{The production of hard particles or jets is strongly suppressed in heavy ion collisions relative to
expectations from incoherent proton-proton interactions}
\item{The dependence of the total multiplicity  scaled by the number of participants on centrality is weak. The total multiplicity of
particles in the forward region is suppressed in deuteron-Gold collisions.. }

\end{itemize}
These discoveries  and more are described in the experimental white papers for the four experiments.\cite{expwhitepaper}

The theoretical implications of these results are strong, and are reviewed in the proceedings of
a Riken Brookhaven Center Workshop, {\it New Discoveries at RHIC}.\cite{newdiscoveries}
The large flow combined with the elliptic flow can be described within hydyrodynamical models
of nuclear collisions.  These hydrodynamical models require somewhat small viscosities corresponding to mean free paths $\lambda \le .5~Fm/c$.  There are uncertainties as to how small a viscosity is allowed
due to uncertainties in the initial conditions for the collision.  Nevertheless, in my opinion, the hydrodynamical descriptions  in the zero mean free path approximation work quite well..  It is quite remarkable that heavy quarks such as charm
flow along with the light quarks, since it is hard to impart a significant flow velocity to a heavy quark.
These observations have led to the paradigm of the Strongly Interacting Quark Gluon Plasma.

The jet energy loss results have shown that out to very large transverse momenta, jets are suppressed relative to incoherent production.\cite{bj}-\cite{glv}  The magnitude of the suppression out to such large $p_T$ was a great surprise.  The apparent suppression of heavy quark jet remains a surprise and is not understood.  The empirical observation of jet quenching allows the jets to explore the
space-time distribution of the produced matter, tomography if you like.\cite{wang}   The strength of the suppression lends support to the idea of the Quark Gluon Plasma as very strongly interacting.

The weak dependence of the multiplicity of produced particles scaled by the multiplicity and the suppression of inclusive particle production in the forward direction of deuteron-Gold collisions was predicted by the theory of the Color Glass Condensate.\cite{newdiscoveries}  Recent results  on particle correlations\cite{srivastava} and on jet correlations\cite{bland} appear to be consistent with this hypothesis.

In the future, a number of new tools experimental tools will be used as probes:

\begin{itemize}
\item{Stochastic beam cooling and increased luminosity:  Charmonium, charm and jet studies}
\item{Detector improvements for dA studies: The Color Glass Condensate}
\item{Low Energy Run: The phase digram at finite density}
\item{Variable energy and variable A and Z: Event by event P and CP vioation}
\end{itemize}
The last item above is described in the talk by Harmen Warringa during this meeting.\cite{warringa}

\section{LHC and Heavy Ions}

At this point I am supposed to predict something about the LHC experiments.  I  quote Mark twain:

{\bf "A prophet doesn't have to have any brains.  They are good to have, of course, but they are of no use in professional work.  It is the restfullest vocation there is."}

\begin{figure}[htbp]
\begin{center}
\includegraphics[width=0.70\textwidth]{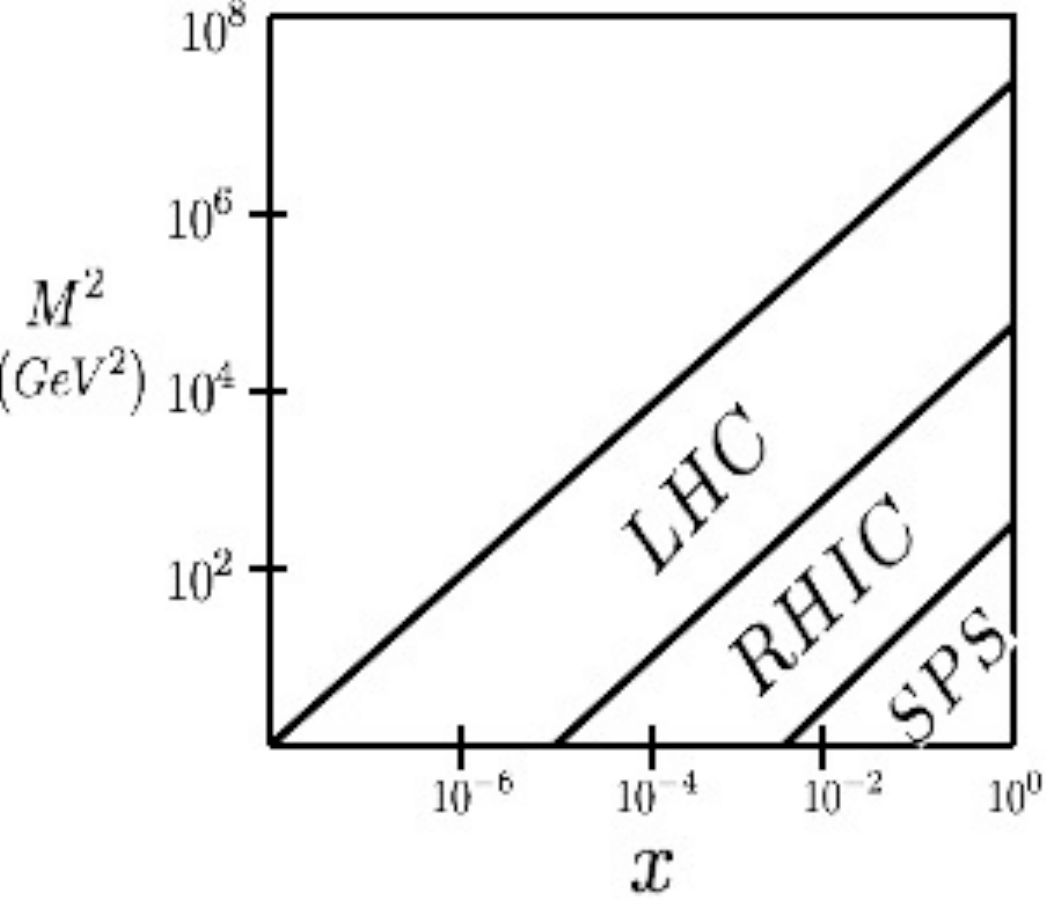}
\end{center}
\caption{A plot of the available phase space for the SPS, RHIC and LHC experiments.
}
\label{lhc}
\end{figure}

In the Figure. \ref{lhc}, I show  a plot of the allowed phase space for SPS, RHIC and LHC in terms
of $ln(1/x)$ and $ln(M^2)$.  Measuring the area of the curves, you conclude that the ratio
of physics excitement is roughly
\begin{equation}
{{LHC} \over {RHIC}} \sim {{RHIC} \over {SPS}}
\end{equation}

Surely the LHC will bring unanticipated discovery.  This will no doubt lead to new ways of thinking,
and may also force us to revise our previous conceptions. 
There is another quote from Mark Twain well worth remembering:

{\bf "It's not what you don't know that gets you in trouble, it's what you think you know."}

\section{Development of Heavy Ion Physics in China and India}

This meeting is in India and the previous one was in China.   Chinese and Indian involvement in our science has resulted in many of the very bright  and creative young and middle career scientists who are shaping our field.

Prof. T. D. Lee has been absolutely essential in developing high quality science in China, and providing a broad framework in which new ideas and effort my be successfully pursued.   The first meeting I know concerning heavy ion physics was in Wuhan, China in 1983.  It was a school organized Lian-Shou Liu.  
Liu had me give eleven lectures, and I remember the meeting very well, and my excitement in being in China for the first time.
Subsequently , Lian-Shou Liu and Ulrich Heinz organized a workshop in 1994.  In 1999, Xin-Nian Wang and Lian-Shou Liu organized a meeting.  These meetings were the direct result of Liu's scientific vision and enthusiasm,  and resulted in Chinese participation in Alice  at LHC and Star at RHIC.  Among the groups involved in heavy ion physics are Central China Normal University,  China Institute of Atomic Energy, and Huazhong University of Science and Technology.  

In India in 1983, Bikash Sinha published Universal Signals of the Quark Gluon Plasma.  The abstract was:

{\it It is shown that the ratio of production rates of photon to muon pairs and pion to muon pairs from a QGP are independent of the space-time evolution of the plasma fireball, and thus are universal signals of the quark-gluon plasma}

This was essentially the beginning of Dr.  Sinha's involvement in the physics of matter at high energy density.  He later went on to work with Dinesh Srivastava on a number of phenomenological issues
related to heavy ion physics.

At about the same time, Rajiv Gavai was at the University of Bielefeld as a young postdoctoral fellow.  he was working with Helmut Satz, and published his first paper on lattice gauge theory.  Rajiv went on to develop a major lattice gauge theory effort in India, which has involved a number of talented and creative Indian physicists.  His work has been in the vanguard of the lattice gauge theory effort at finite temperature and density, and in advanced scientific computing in India.

Dr. Sinha's involvement in heavy ion physics led to a major involvement in the experimental programs at the SPS, RHIC and Alice.  The Indian groups have had the major responsibility for the Photon Multiplicity Detector at STAR and ALICE.  Bikash also began the series of meeting in India:  The Physics and Astrophysics of the Quark Gluon Plasma, first held in 1988.  These meetings brought together the international scientific community with the young developing  Indian community.  I have attended many of these meetings, and I have always learned something new at each of them, both scientific and cultural.

\section{Summary}

The heavy ion effort is about scientific discovery.  I finish the talk with a quote from {\it Innocents Abroad} by Mark Twain:

{\bf  "What is it that confers the noblest delight?  What is it which swells a man's breast with pride above that which any other experience can give him?  Discovery!  To know that you are walking where none others have walked; that you are beholding what no human eye has seen before; that you are breathing a virgin atmosphere.  To give birth to an idea - an intellectual nugget - right under the dust of a field that many a brain plow had gone over before.  To be first - that is the idea.  To do something, say something, see something before anybody else- these are the things that confer a pleasure compared with other pleasures are tame and commonplace, other ecstasies cheap and trivial.  Lifetimes of ecstasy crowded into a single moment."}

\section*{Acknowledgments}
I thank my colleagues at BNL and RBRC-BNL Center for their kind advice in presenting
this talk.  I thank Nu Xu and Bikash Sinha for providing me information on the
Chinese and Indian collaboration in the BNL and CERN heav ion programs. 
My research is supported under DOE Contract No.
DE-AC02-98CH10886.

\end{document}